\documentclass[
reprint,
preprintnumbers,
nofootinbib,
amsmath,
amssymb,
aps,
prd
]{revtex4}
\usepackage{enumerate}
\usepackage{color}
\usepackage{graphicx}
\usepackage{dcolumn}
\usepackage{bm}
\usepackage{empheq}
\usepackage{nomencl}
\usepackage{hyperref}
\usepackage{mathrsfs}
\usepackage{amsmath}
\usepackage{epstopdf}
\hypersetup
{
	colorlinks,%
	citecolor=blue,%
	linkcolor=magenta,%
	urlcolor=blue,%
}

\usepackage{subfigure}

\begin{document}
	\newcommand{\be}{\begin{equation}}
\newcommand{\en}{\end{equation}}
 \newcommand{\bea}{\begin{eqnarray}}
 \newcommand{\ena}{\end{eqnarray}}
  \newcommand{\sch}{Schwarzschild}
  \newcommand{\wdt}{\widetilde}
  \newcommand{\red}[1]{\textcolor{red}{#1}}
	\newcommand{\green}[1]{\textcolor{green}{#1}}
	\newcommand{\blue}[1]{\textcolor{blue}{#1}}
	\newcommand{\cyan}[1]{\textcolor{cyan}{#1}}
	\newcommand{\purple}[1]{\textcolor{purple}{#1}}
	\newcommand{\yellowbox}[1]{\colorbox{yellow}{#1}}
	\newcommand{\purplebox}[1]{\colorbox{purple}{#1}}
	\newcommand{\yellow}[1]{\textcolor{yellow!70!red}{#1}}
	\title{Dirac clouds around dilatonic black holes}
	\author{Yang Huang $^1$}\email{sps\_huangy@ujn.edu.cn}
	\author{Hongsheng Zhang $^{1}$}\email{corresponding author: sps\_zhanghs@ujn.edu.cn}
	\affiliation{
		$^1$ School of Physics and Technology, University of Jinan, 336, West Road of Nan Xinzhuang, Jinan 250022, Shandong, China}
	
	\begin{abstract}
	Dirac cloud is in absence in general relativity since the superradiance mechanism fails to work for Dirac fields. For the first time we find a novel mechanism to support Dirac clouds, which is independent on superradiance
     mechanism. We    study quasi bound states of Dirac particles around a charged spherical black hole in dilatonic gravity. We find that the quasi bound states become real bound states when the central black hole becomes extremal.    We make an intensive study of the energy spectrum of the stationary clouds for different fine structure constant $\mu M$, and reveal the existence condition of these clouds. Our result strongly implies that extreme dilatonic black holes behave as elementary particles.
	\end{abstract}
	\maketitle
	


 The black hole uniqueness theorem is a milestone in black hole physics, which implies that numerous characters, such as multi-poles, baryon number, lepton number, chemical composition, etc, of the progenitor
 will be dissipated in the formation process of black holes except mass, electric charge, and angular momentum in Einstein-Maxwell theory \cite{Bekenstein:1972ny}. Since the proposition of the uniqueness theorem, explorations of new hair
 of black holes never cease. To find an analytical metric of a hairy black hole is certainly admirable, but extremely difficult. Back to general tradition in physics, a feasible approach is to
  introduce a small perturbation around a black hole, and then to observe the evolution of the small perturbation. If the perturbation can enter a stationary state, it becomes a cloud of the hole, just like an electron cloud
  in a hydrogen atom. If one further considers the back reaction of the cloud to the metric, one may reach a hairy black hole \cite{Herdeiro:2014goa,Hong:2020miv}.

  The most important and practicable road to develop a cloud in astrophysics depends on superradiance mechanism.  The superradiance mechanism is an extension of the Klein paradox, which says that more particles will be excited when a scalar particle scattered by a special potential. Physically, the collision of the particle and the potential leads to pair production and the anti particle with negative energy is absorbed by the potential. For a rotating (Kerr) black hole, a scalar particle passing through the ergosphere, i.e.,  scattered by the effective potential of the ergosphere of the hole, may generate more particles sensed by observers at spacelike infinity. This is the black hole superradiance mechanism. There is a complementary effect of superradiance, which reduces the particle number around a black hole. This may be dubbed super absorption effect, which
  indicates that any surrounding particles of a black hole has a probability to permeate into the hole because of quantum tunneling effect. If the superradiance and super absorption effects get a fine balance, the life time of surrounding particle of a black hole can be very long, even longer than the age of the universe. In this case, the  surrounding particles compose a cloud of a black hole \cite{Huang:2016qnk,Huang:2017whw,Hod:2012px,Bernard:2016wqo}.

  A point worth emphasizing is that the superradiance mechanism works only for Boson. Because of Pauli's exclusion principle, Fermion never undergoes  superradiance process. While the super absorption mechanism still works for a Fermion, and has no complementary mechanism to balance it. The conclusion is that Dirac particles cannot develop a cloud around a black hole since they inevitably infiltrate into the hole, and disappear soon. Fermions are fundamental bricks of the Universe. It is quite significative to find novel mechanism to develop a Dirac cloud for a black hole.

  We will show that a stationary Dirac cloud around a GMGHS (Gibbons-Maeda-Garfinkle-Horowitz-Strominger) black hole is able to survive when the hole becomes an extreme one. The GMGHS solution in dilatonic gravity is an influential solution, which is originally presented as a black hole in low energy heterotic string theory \cite{Gibbons:1987ps,Garfinkle:1990qj}. Dilatonic gravity (scalar-tensor theory) is a historic and alive theory of modified gravities, which can be traced back to Kaluza-Klein compactification and Dirac's large number hypothesis. Many different modified gravities reduce to dilatonic gravity at low energy limit.

The dilatonic black hole we consider here is described by the GMGHS metric,
\begin{equation}\label{Eq: metric}
	ds^2=-Fdt^2+F^{-1}dr^2+r^2G\left(d\vartheta^2+\sin^2\vartheta d\varphi^2\right),
\end{equation}
with $F(r)=1-r_+/r$ and $G(r)=1-r_-/r$, where $r_+=2M$, $r_-=Q^2/M$. The black hole mass and charge are given by $M$, and $Q$, respectively. It is useful to parameterize the black hole charge by $q=Q/Q_{\text{max}}=1-e^{-\eta}$. The black hole becomes extremal when $\eta\rightarrow\infty$.

We first consider the quasibound states of massive Dirac fields in GMGHS spacetimes. The equation of a massive Dirac field in curved spacetime is written as
\begin{equation}
	\left[\gamma^\alpha\left(\partial_\alpha-\Gamma_\alpha\right)-\mu\right]\bm{\Psi}=0,
\end{equation}
where $\mu$, $\bm{\Psi}$, and $\Gamma_\alpha$ are the mass, the Dirac-four spinor, and spinor connection matrices, respectively.
A suitable choice of matrices $\gamma^\mu$ obeying the relation $\gamma^\mu\gamma^\nu+\gamma^\nu\gamma^\mu=2g^{\mu\nu}_{\text{dilaton}}$ are given by $\gamma^\mu=e^{\mu}_{a}\hat{\gamma}^a$, where $e^{\mu}_{a}$ are tetrad fields and matrices $\hat{\gamma}^a$ are defined by the flat space Dirac matrices $\tilde{\gamma}^a$, i.e., $\hat{\gamma}^0=i\tilde{\gamma}^0$, $\hat{\gamma}^1=i\tilde{\gamma}^3$, $\hat{\gamma}^2=i\tilde{\gamma}^1$, and $\hat{\gamma}^3=i\tilde{\gamma}^2$, with
\begin{equation}
\tilde{\gamma}^0=
\left(
\begin{array}{c c}
0   &  I_2\\
I_2 & 0
\end{array}
\right),\;\;\;
\tilde{\gamma}^j=
\left(
\begin{array}{c c}
0         & \sigma_j\\
-\sigma_j & 0
\end{array}
\right).
\end{equation}
The spinor connection matrices are written as
\begin{equation}
	\Gamma_\alpha=-\frac{1}{4}g_{\mu\nu}e^{\mu}_{a}e^{\nu}_{b;\alpha}\hat{\gamma}^{a}\hat{\gamma}^{b}.
\end{equation}
Using the metric (\ref{Eq: metric}), we find the explicit form of the components of $\Gamma_\alpha$
\begin{equation}
\begin{aligned}
\Gamma_t&=\frac{F'}{4}\hat{\gamma}^{0}\hat{\gamma}^{1},\;\;\;\;\;\;\;\Gamma_r=0,\\
\Gamma_\vartheta&=W\;\hat{\gamma}^{1}\hat{\gamma}^{2},\\
\Gamma_\varphi&=W\sin\vartheta\;\hat{\gamma}^{1}\hat{\gamma}^{3}+\frac{\cos\vartheta}{2}\hat{\gamma}^{2}\hat{\gamma}^{3},
\end{aligned}
\end{equation}
where
\begin{equation}
W(r)=\frac{r}{4}\sqrt{\frac{F}{G}}\left(\frac{2}{r}+\frac{d}{dr}\right)G.
\end{equation}
Defining $\bm{\Phi}=rG^{1/2}F^{1/4}\left(\sin\vartheta\right)^{1/2}\bm{\Psi}$, we can rewrite the Dirac equation as \cite{RevModPhys.29.465}
\begin{equation}\label{Eq: Schrodinger}
i\partial_t\bm{\Phi}=H\bm{\Phi},
\end{equation}
where
\begin{equation}
\begin{aligned}
H&=i\hat{\gamma}^0\hat{\gamma}^1F\partial_r+\frac{1}{r}\sqrt{\frac{F}{G}}\hat{\gamma}^1K-i\mu\sqrt{F}\hat{\gamma}^0,\;\;\text{and}\\
K&=i\hat{\gamma}^1\hat{\gamma}^0\hat{\gamma}^2\partial_\vartheta+i\hat{\gamma}^1\hat{\gamma}^0\hat{\gamma}^3\csc\vartheta\partial_\varphi.
\end{aligned}
\end{equation}
The angular operator $K$ is Hermitian and commutes with $H$, $[K,H]=0$. Therefore, operators $H$ and $K$ have simultaneous eigenfunctions, whose $\mu$-th component can be separated by
\begin{equation}\label{Eq: ansatz}
\Phi_{\mu}=R_{\mu}(r)\Theta_{\mu}(\vartheta,\varphi)e^{-i\omega t}.
\end{equation}
The angular part $\Theta_{\mu}(\vartheta,\varphi)$ is determined by the requirement $K\bm{\Phi}=\lambda\bm{\Phi}$, where $\lambda$ is constant. The orbital angular momentum and spin angular momentum are no longer conserved quantities, respectively. $\lambda$ denotes the quantum number of the total angular momentum. Boundary conditions at $\vartheta=0,$ and $\pi$ determine a spectrum of non-zero integers of $\lambda$, i.e., $\lambda=\pm1,\pm2,\cdots$ \cite{RevModPhys.29.465}.

Substituting the ansatz (\ref{Eq: ansatz}) into Eq.(\ref{Eq: Schrodinger}), and replacing the operator $K$ by the number $\lambda$, we obtain $H\bm{\Phi}=\omega\bm{\Phi}$. After replacing the operator $K$ by number $\lambda$, only two matrices $\hat{\gamma}^0$ and $\hat{\gamma}^1$ remain explicitly in $H$. The radial factor can therefore be represented by a two component spinor
\begin{equation}
R=\left(
\begin{array}{c}
R_+\\
R_-
\end{array} \right).
\end{equation}
In this representation, the radial equations are given by
\begin{equation}\label{Eq: radial eq}
	\mathscr{D}_{\pm}R_{\pm}=\mathscr{C}_{\mp}R_{\mp},
\end{equation}
where $\mathscr{D}_{\pm}=F\partial_r\pm i\omega$ and
\begin{equation}\label{Eq: Def of Cpm}
	\mathscr{C}_{\pm}=\frac{1}{r}\sqrt{\frac{F}{G}}\left(\lambda\pm i\mu r\sqrt{G}\right).
\end{equation}
It is very clear that the above two equations reduce to the Dirac equation of the case of \sch\ when $G=1$ \cite{Lasenby:2002mc}. One can eliminate one of the two dependent variables to obtain a single second order ordinary differential equation for the other variable.

Generally, quasibound states are purely ingoing at the event horizon and vanish at spatial infinity.
Our aim is to compute the quasibound states of massive Dirac fields in non-extreme GMGHS spacetime. Close to $r=r_+$, the ingoing wave solutions of Eqs.(\ref{Eq: radial eq}) are given by
\begin{equation}\label{Eq: horizon sol}
	R_{\pm}\sim\mathcal{T}_{\pm}(r-r_+)^{\frac{1\mp1}{4}}e^{-i\omega r_*},
\end{equation}
where $r_*=\int dr/F$ is the tortoise coordinate, and $\mathcal{T}_{\pm}$ are the transmission coefficients. At infinity $r\rightarrow\infty$, the vanishing solutions are given by
\begin{equation}\label{Eq: far region sol}
	R_{\pm}\sim r^{\chi}e^{kr},
\end{equation}
where $k=-\sqrt{\mu^2-\omega^2}$, and $\chi=M(\mu^2-2\omega^2)/k$. With these boundary conditions, we solve the second order equation of $R_+$ derived from Eqs.(\ref{Eq: radial eq}) to obtain the quasibound state frequencies. To get high accurate eigen-frequency of the quasibound state, the spectral method is used to solve this eigenvalue problem.

For bounded modes with $\omega<\mu$, the field cannot propagate to spatial infinity. However, the system is still nonconservative, since the field will leaks into the black hole through the horizon. Therefore, the quasibound state frequencies are in general complex, i.e., $\omega=\omega_R+i\omega_I$. We find that $\omega_I$ is always negative for Dirac fields in non-extreme GMGHS spacetime, which excludes the existence of stationary states in this case. Physically, negative $\omega_I$ implies that the Dirac particles have probabilities to permeate into the black hole.

\begin{figure*}
	\centering	
	\includegraphics[width=0.45\textwidth,height=0.35\textwidth]{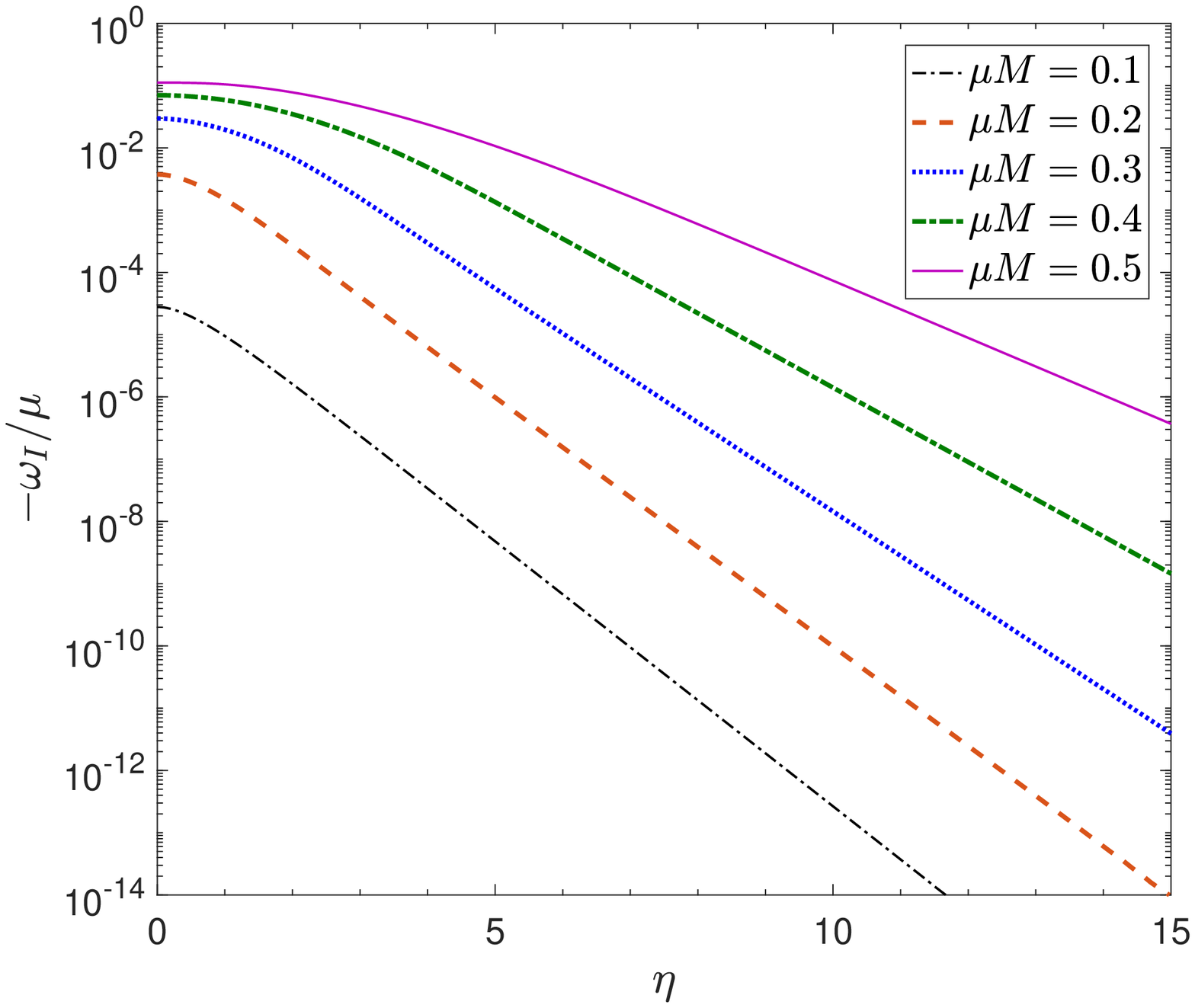}	
	\includegraphics[width=0.45\textwidth,height=0.35\textwidth]{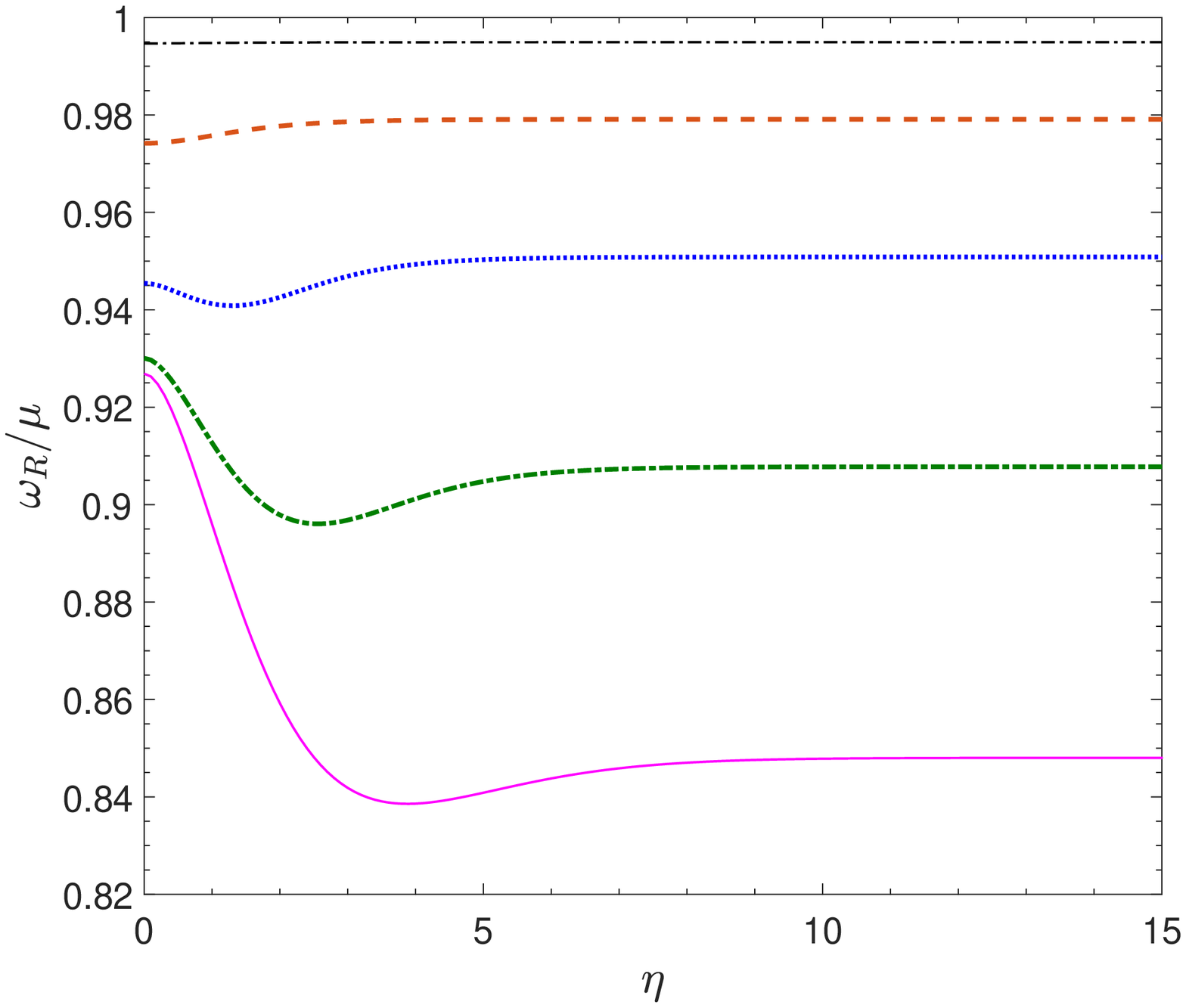}
	\caption{Eigen frequencies of quasibound state, with $\lambda=-1$, as functions of $\eta$ for different values of $\mu M$. The left panel shows the imaginary parts, whereas the right panel shows the real part.}
	\label{Fig: quasibound states}
\end{figure*}

Figure \ref{Fig: quasibound states} shows the typical numerical results of the quasibound state frequencies as functions of $\eta$. We see that both $\omega_R$ and $\omega_I$ are sensitive to the values of $\eta$, or the black hole charge. In particular, when $\eta\gg1$, $\omega_R$ tends to constant, and $|\omega_I|$ decays to zero exponentially. This result is in analogy with the case of a massive scalar field in non-extreme GMGHS spacetime \cite{Huang:2020pga}. It was shown that in the extreme limit $\eta\rightarrow\infty$, the potential barrier becomes infinitely wide, and the tunneling effect could be completely suppressed. Therefore, real bound states with $\omega_I=0$ may exist for massive fields in extreme GMGHS spacetime.
  \begin{figure}
	\centering	
	\includegraphics[width=0.45\textwidth,height=0.35\textwidth]{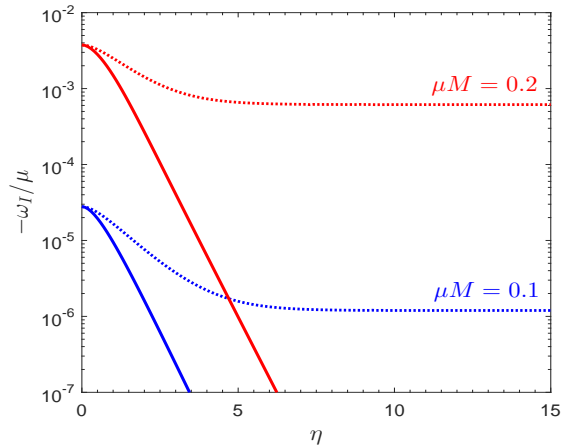}
	\caption{Dotted curves denote the the imaginary part of eigen frequencies of Dirac particles in the case of RN, while solid curves for the case of GMGHS. We set $\lambda=-1$.}
	\label{CRN}
\end{figure}

As a comparison, in Fig.\ref{CRN}, we plot the imaginary part of eigen frequencies of quasibound state of Dirac particles on the back ground of a Reissner-Nordstrom black hole, which is the corresponding one of GMGHS in general relativity. Note that we have parameterized the RN charge in the same way, i.e., $Q_{\text{RN}}/Q_{\text{RN, max}}=1-e^{-\eta}$. From Fig \ref{CRN}, it is clear that Dirac particles permeate into an extreme RN hole since the imaginary part of the eigen frequencies go to constant when the hole becomes extremal, while the permeating process will cease for an extreme GMGHS hole.

With the above observations, we have every reason to conjecture that an extreme GMGHS black hole permits real bound state of Dirac particles. To investigate the existence of real bound states of massive Dirac field, we solve the Dirac equation in extreme GMGHS spacetime, in which $F=G=1-r_+/r$, and $\mathscr{C}_{\pm}=(\lambda\pm i\mu r\sqrt{F})/r$, see Eq.(\ref{Eq: Def of Cpm}). First, we eliminate $R_-$ in Eq.(\ref{Eq: radial eq}) to obtain the second order equation of $R_+$, and introduce a new dimensionless coordinate
\begin{equation}
	x\equiv\frac{\sqrt{(r-r_+)\mu}}{\sqrt{(r-r_+)\mu}+1}\in\left[0,1\right].
\end{equation}
The black hole horizon is reached when $x=0$, whereas the infinity is denoted by $x=1$. Then, the second order equation of $R_+$ can be written as
\begin{equation}\label{Eq: 2nd eq R+}
	\left(\frac{d^2}{dx^2}+\mathcal{C}_1\frac{d}{dx}+\mathcal{C}_2\right)R_+=0,
\end{equation}
where the form of $\mathcal{C}_i$ are too cumbersome to be presented here. One can also eliminate $R_+$ to obtain the second order equation of $R_-$, but the final result remains unchanged.

\begin{figure}
	\centering	
	\includegraphics[width=0.45\textwidth,height=0.35\textwidth]{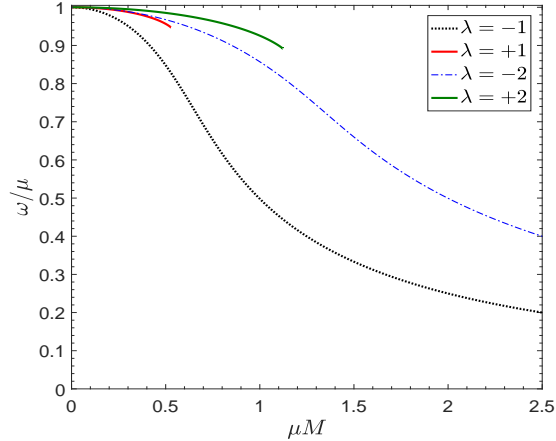}		
	\caption{Frequency of the ground states as functions of $\mu M$, for different values of $\lambda$.}
	\label{Fig: cloud}
\end{figure}

\begin{figure}
	\centering	
	\includegraphics[width=0.45\textwidth,height=0.35\textwidth]{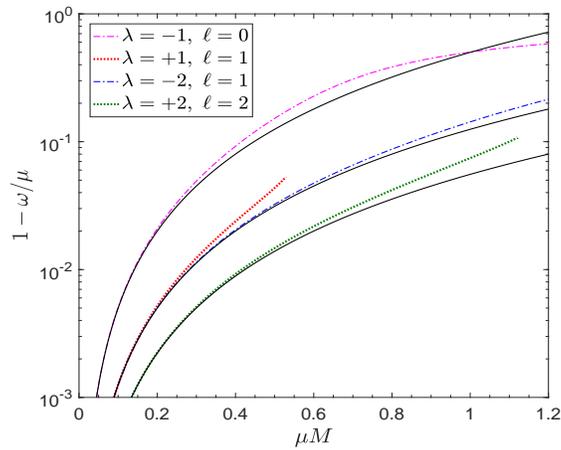}	
	\caption{Binding energy of the node-less ($n=0$) states for different values of $\lambda$. The solid curves are given by Eq.(\ref{Eq: analytic w}).}
	\label{Fig: binding energy}
\end{figure}

\begin{figure}
	\centering	
	\includegraphics[width=0.45\textwidth,height=0.35\textwidth]{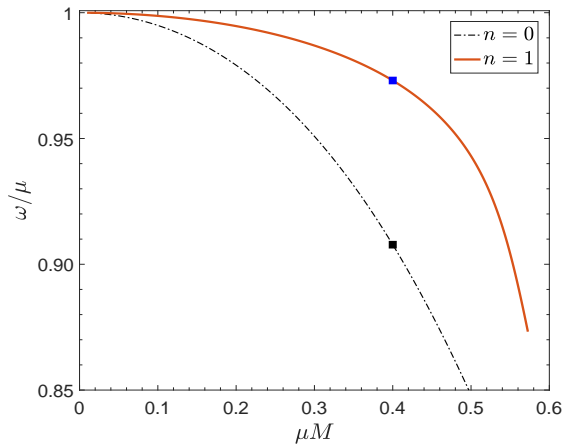}	
	\caption{Comparison of the bound state frequencies of the first two excited states, with $\lambda=-1$. The two points mark the states for $\mu M=0.4$, whose radial functions are presented in Fig. \ref{Fig: radial func}.}
	\label{Fig: cloudn1}
\end{figure}

\begin{figure}
	\centering		
	\includegraphics[width=0.45\textwidth,height=0.35\textwidth]{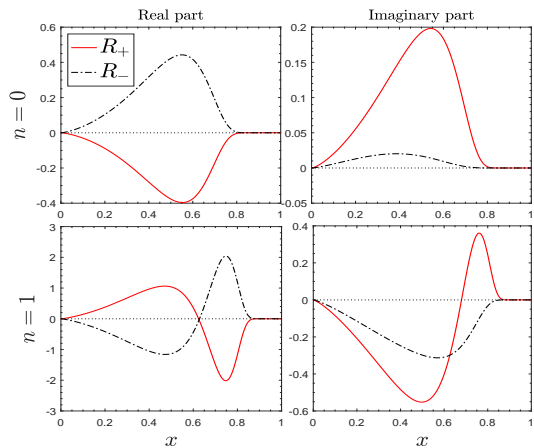}		
	\caption{Radial profiles of the bound states marked in Fig. \ref{Fig: cloudn1}.}
	\label{Fig: radial func}
\end{figure}

To obtain the proper boundary conditions, we first investigate the asymptotic behavior of the Dirac equation in the background of an extreme hole. At infinity, the asymptotic solution of Eq.(\ref{Eq: 2nd eq R+}) is given by Eq.(\ref{Eq: far region sol}). However,  the ingoing waves at the horizon (\ref{Eq: horizon sol}) are no longer asymptotic solutions of the radial equation. Instead, the exact regular solutions (with $M\omega\le|\lambda|/2$) at this boundary are now given by
\begin{equation}\label{Eq: horizon sol ex}
	R_{+}\sim x^{2\alpha},
\end{equation}
where $\alpha=\sqrt{\lambda^2-4M^2\omega^2}$. We seek the solutions of Eq.(\ref{Eq: 2nd eq R+}) obeying boundary conditions (\ref{Eq: far region sol}) and (\ref{Eq: horizon sol ex}). This is an eigenvalue problem of $\omega$. We apply the spectral method together with the Newton-Raphson method \cite{Dias:2015nua}. This enables us to determine the eigenvalue of $\omega$ and the corresponding radial function simultaneously. After fixing $\lambda$, we find a series of solutions by sweeping the $\mu$ space. The numerical results are presented in Figs. \ref{Fig: cloud}-\ref{Fig: radial func}.

In Fig.\ref{Fig: cloud}, we plot the frequency of the ground states for different values of $\lambda$. As expected, these frequencies are purely real, and they are exactly the extreme limit of the quasibound state frequencies in the non-extreme case. Thus, the solutions we have found are real bound states, as they will not grow or decay with time. Thus the particles in bound state develop a stationary cloud over the black hole horizon.

More interestingly, when $\mu M\ll1$, the real bound state we have found has a hydrogen-like spectrum \cite{Ternov:1980st,Huang:2017nho}
\begin{equation}\label{Eq: analytic w}
	\omega\approx\mu\left(1-\frac{\mu^2M^2}{2\tilde{n}^2}\right),\;\;\;\tilde{n}=n+\ell+1,
\end{equation}
where $n=0,1,2,\cdots$, is the node number of the radial functions, whereas $\ell=|\lambda+1/2|-1/2=0,1,2\cdots$, is the orbital angular momentum. The energy level is determined by the principle quantum number $\tilde{n}=1,2,3,\cdots$. The consistency between Eq.(\ref{Eq: analytic w}) and numerical results is presented in Fig. \ref{Fig: binding energy}.

Except for ground state, we further explore the excited bound state. Figure \ref{Fig: cloudn1} shows the bound state frequencies of the first two excited states. A common feature of Figs. \ref{Fig: cloud} and \ref{Fig: cloudn1} is that some of the curves have endpoints on the right side, which implies that no bound states exist when $\mu M$ is greater than some critical value. In fact, we find that approaching to the critical value of $\mu M$, parameter $\alpha$ defined in Eq.(\ref{Eq: horizon sol ex}) tends to zero, as such, the regularity of the radial function at the horizon becomes delicate.

In Fig. \ref{Fig: radial func}, we show the radial functions of the bound states marked in Fig. \ref{Fig: cloudn1}. Although the eigenvalue of $\omega$ is purely real, the corresponding radial functions are in general complex. As expected, we see that all the radial functions, both the real and imaginary parts, are smooth and finite in the whole range of $x$. Thus, we conclude that the extreme GMGHS can support stationary Dirac clouds.

\begin{figure*}
	\centering		
	\includegraphics[width=0.45\textwidth,height=0.33\textwidth]{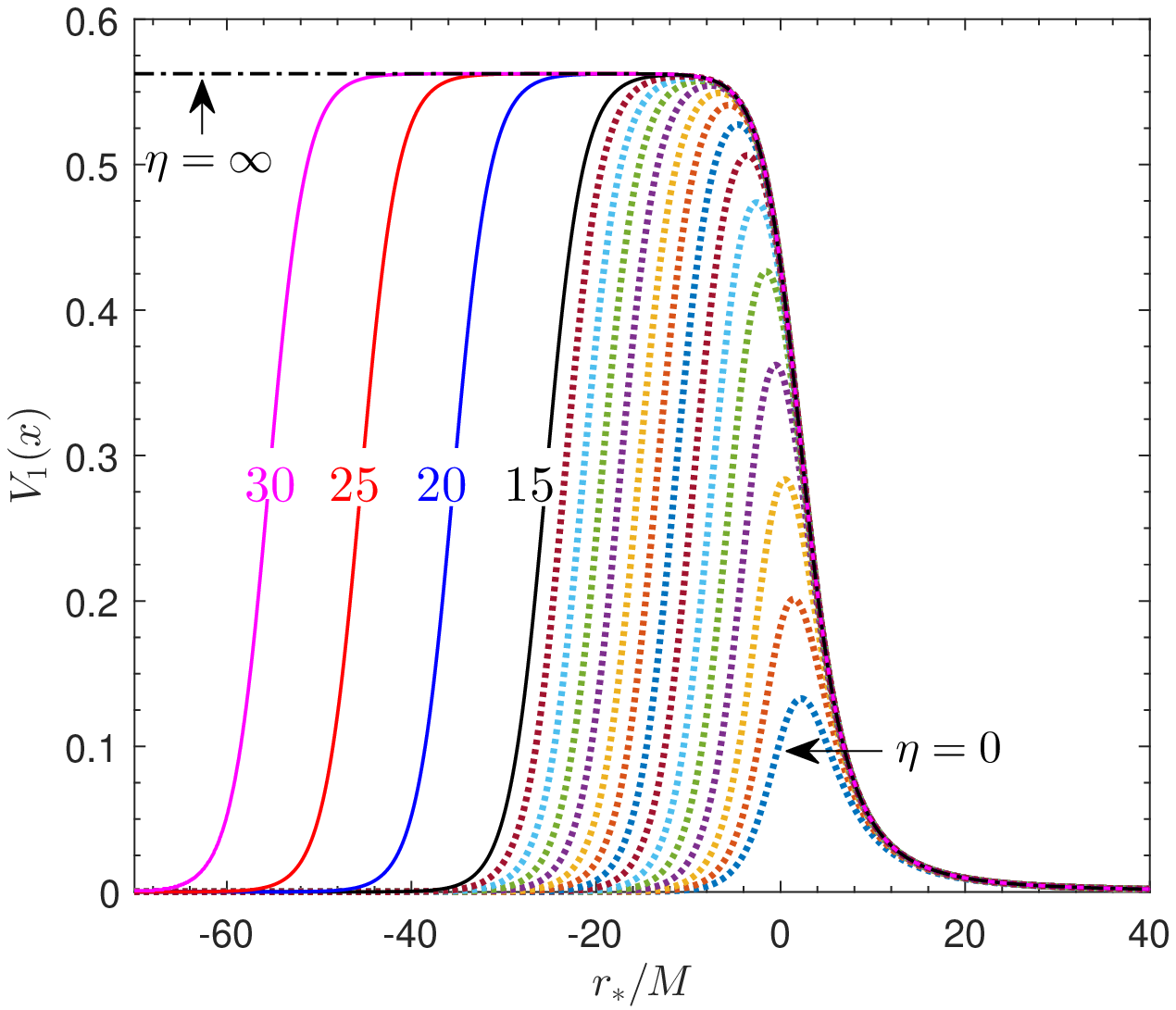}
	\includegraphics[width=0.45\textwidth,height=0.33\textwidth]{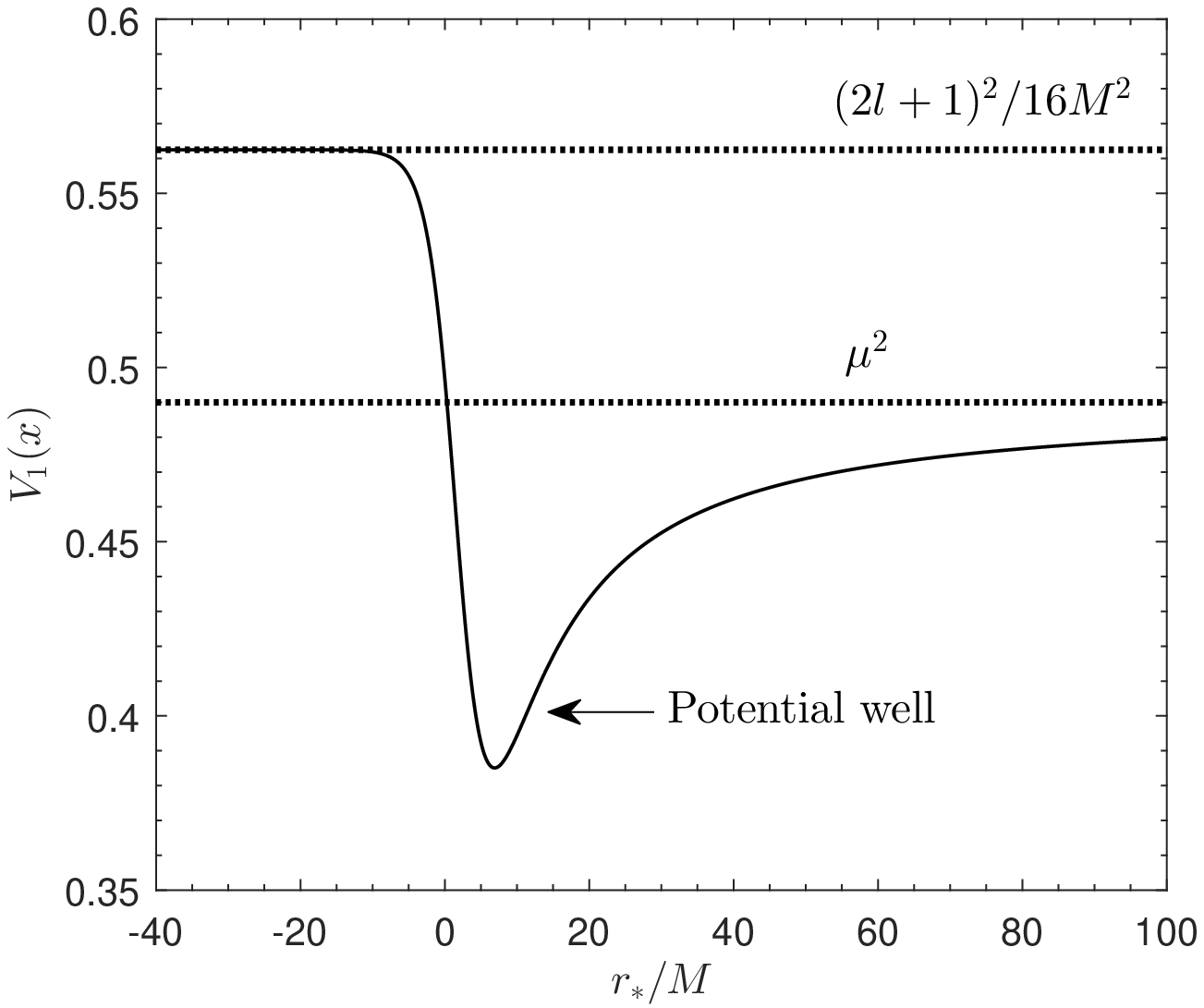}
	\caption{Effective potential (with $l=1$) of the radial equation of massive scalar field in GMGHS spacetime. In the left panel, we fix the mass $\mu=0$, and compare the potential for different values of $\eta$. The dotted curves correspond to $\eta=0,1,2,\cdots,14$, and the solid curves correspond to $\eta=15,20,25,30$. In the right panel, we show the effective potential for $\eta=\infty$, and the mass of the scalar field is $\mu M=0.7$.}
	\label{Fig: potential}
\end{figure*}

 We present a detailed analysis of the mechanism for this Dirac cloud. We have found that extremal GMGHS black hole can also support stationary scalar clouds in its exterior spacetime \cite{Huang:2020pga}. The mechanism for Dirac clouds in GMGHS black hole spacetime is similar to the scalar field. First, we illustrate the mechanism for the scalar clouds. After separation of variables, the radial function of the scalar field is given by
\begin{equation}\label{Eq: the radial eq}
	\frac{d^2\psi}{dr^2_*}+\left[\omega^2-V_l(r)\right]\psi=0,
\end{equation}
with $r_*=\int dr/F$ the tortoise coordinate, and the effective potential given by
\begin{equation}\label{Eq: effective potential}
	V_l(r)=\frac{F(r)}{G(r)}\left[\frac{F'(r)}{r}+\frac{l(l+1)}{r^2}+\mu^2G(r)\right]-\frac{2M^2q^2}{r^4}\frac{F(r)}{G(r)^2}\left[1+\frac{q^2}{2}\left(1-\frac{6M}{r}\right)\right].
\end{equation}
We plot the potential as functions of $r_*$ in Fig. \ref{Fig: potential}. In the left panel, we see that the width of the potential barrier increases monotonously with the increase of $\eta$. In the extreme limit $\eta\rightarrow\infty$, the width of the potential barrier increases without bound, whereas the height tends to $\omega^2_0\equiv(2l+1)^2/16M^2$. In this case, the potential (\ref{Eq: effective potential}) can be written as
\begin{equation}
	V_l(r)=\frac{l(l+1)}{r^2}+\frac{M}{r^3}\left(2-\frac{3M}{r}\right)+F(r)\mu^2.
\end{equation}
One may check that $V_{l}(r\rightarrow 2M)=\omega^2_0$.
This implies that a low frequency mode with $\omega<\omega_0$ coming from infinity will be totally reflected back to infinity by the potential barrier. If the scalar field has a nonzero mass ($\mu\neq0$), a potential well appears at $r_*/M\approx0$, as is shown in the right panel  of the of fig. \ref{Fig: potential}.  Since both sides of the potential well is infinitely wide, real bound states with $\omega_I=0$ does exist. Clearly, the bound state frequencies are purely real, and must obey $\omega<\text{min}\left\lbrace \omega_0,\mu\right\rbrace$.

In obtaining the bound state solutions, we choose boundary conditions as follows. At infinity, the potential tends to $\mu^2$, and the most general solution of the radial equation (\ref{Eq: the radial eq}) is given by
\begin{equation}
	\psi= C_1e^{-\sqrt{\mu^2-\omega^2}r_*}+C_2e^{\sqrt{\mu^2-\omega^2}r_*}.
\end{equation}
For bound states, we have $C_2=0$. At the horizon, $V_{l}(r\rightarrow 2M)=\omega^2_0$, and the most general solution is 
\begin{equation}
	\psi= B_1e^{-\sqrt{\omega_0^2-\omega^2}r_*}+B_2e^{\sqrt{\omega_0^2-\omega^2}r_*}.
\end{equation}
Close to the horizon $r_*\rightarrow-\infty$, we have $B_1=0$ for bound states. Thus, the near horizon behavior of the radial function is
\begin{equation}\label{Eq: BC at horizon}
	\psi(r\rightarrow 2M)\sim e^{\sqrt{\omega_0^2-\omega^2}r_*}\sim (r-2M)^{\beta},
\end{equation}
where
\begin{equation}\label{Eq: def of beta}
	\beta=2M\sqrt{\omega_0^2-\omega^2}=\sqrt{\left(l+\frac{1}{2}\right)^2-4M^2\omega^2}.
\end{equation}
Clearly, for bound states with $\omega<\omega_0$, $\beta$ is purely real, and the radial function is vanishing at the horizon. Note that Eq.(\ref{Eq: BC at horizon}) does not represent a purely ingoing wave at the horizon for $\omega<\omega_0$. As discussed above, when the GMGHS black hole is extremal, the potential does not vanish at the horizon, and the ingoing wave is no longer an asymptotic solution of the radial equation. Imposing vanishing boundary conditions on both sides, we find the eigenfrequencies of the scalar cloud
\begin{equation}
	\omega=\mu\left[1-\frac{\mu^2 M^2}{2\tilde{n}^2}+\mu^4M^4\left(\frac{15}{8\tilde{n}^4}-\frac{2}{\tilde{n}^3L}\right)+\mathcal{O}\left(\mu^6M^6\right)\right],
\end{equation}
where $L=l+1/2$, and $\tilde{n}=n+l+1$. The frequency is purely real, and is exactly the extreme limit of the quasibound state frequency in the non-extreme case \cite{Huang:2020pga}.


In summary, the existence of scalar clouds around the extremal GMGHS black hole is resulting from a non vanishing potential at the horizon. Such solutions does not exist in an extremal RN black hole spacetime, since the effective potential is always vanishing in the RN case.

Let us now discuss the mechanism for Dirac clouds. We first eliminate $R_-$ to obtain the second order equation of $R_+$. Then, by introducing a new radial function $H_+=\mathscr{C}^{1/2}_{-}R_+$, we rewrite the second order equation in the Schr\"{o}dinger-like form
\begin{equation}\label{Eq: radial eq for H}
	\frac{d^2}{dr^2_*}H_++\left(\omega^2-V_+\right)H_+=0,
\end{equation}
where
\begin{equation}
	V_+=\mathscr{C}_{+}\mathscr{C}_{-}+\frac{3}{4}\left(\frac{\mathscr{C}'_{-}}{\mathscr{C}_{-}}\right)^2-\frac{\mathscr{C}''_{-}-2i\omega \mathscr{C}'_{-}}{2\mathscr{C}_{-}},
\end{equation}
and $\mathscr{C}_{\pm}$ are defined in Eq.(\ref{Eq: Def of Cpm}). Here, prime denotes the derivative with respect to $r_*$. Note that the effective potential of the radial equation is not real. For this reason, we cannot exhibit the potential well as before. However, the potential behaves similarly at at the horizon, i.e., when the GMGHS becomes extremal, the potential does not vanish at $r=r_+$. We find that $V_+(r\rightarrow r_+)=\lambda^2/4M^2=(2j+1)^2/16M^2$, for $\eta=\infty$ ($q=1$), where $j$ is the total angular momentum. Therefore, the most general solution of the radial equation (\ref{Eq: radial eq for H}) at $r=r_+$ can be written as
\begin{equation}
	H_+= B_1e^{-\alpha r_*}+B_2e^{\alpha r_*},
\end{equation}
where
\begin{equation}
	\alpha=\sqrt{\lambda^2-4M^2\omega^2}=\sqrt{\left(j+\frac{1}{2}\right)^2-4M^2\omega^2}.
\end{equation}
One can see that this is very similar to Eq.(\ref{Eq: def of beta}) for scalar clouds. Again, we have $B_1=0$ for bound states. Thus, the radial function of the bound state obeys the boundary condition
\begin{equation}
	R_+\sim e^{\alpha r_*}\sim (r-2M)^{\alpha}\sim x^{2\alpha},
\end{equation}
at the horizon [See Eq.(\ref{Eq: horizon sol ex})]. Although the mechanism for Dirac clouds is not as obvious as that of the scalar clouds, it is a natural extension of the latter one.

Now we make a concise summary of this Letter. To find a cloud is a key step to find the hair of a black hole. Generally, the existence of cloud depends on superradiance mechanism, while for Dirac particles such a mechanism does not work. We demonstrate that Dirac cloud exists over an extreme GMGHS black hole. We first investigate the quasi bound states of Dirac particles around a GMGHS black hole. Then we show that the quasi bound states will be real bound ones when the hole becomes extremal. We then directly solve the Dirac equation in the background of an exactly extreme GMGHS hole, and find bound states of the Dirac particles. These bound states are stationary, and in particular, they are exactly the extreme limit of the quasibound states in non-extreme backgrounds.

The other interesting point related to our results is the proposition that extreme dilatonic black hole behaves like an elementary particle.  The entropy of extreme GMGHS black hole is zero, which implies that it may has no internal freedoms, or its internal freedom can never detected by an observer outside the horizon. This property leads to a conjecture such a hole may behave follow an elementary particle. Our results demonstrate that an external Dirac cloud around a GMGHS black hole never deacy, which presents intense evidence that extreme dilatonic black holes behave as elementary particles \cite{Holzhey:1991bx}.

\begin{acknowledgments}
	This work is supported by the National Key Research and Development Program of China (No. 2020YFC2201400), as well as Shandong Province Natural Science Foundation under grant No. ZR201709220395.
\end{acknowledgments}

\bibliography{cloud}

\begin{thebibliography}{16}
\expandafter\ifx\csname natexlab\endcsname\relax\def\natexlab#1{#1}\fi
\expandafter\ifx\csname bibnamefont\endcsname\relax
  \def\bibnamefont#1{#1}\fi
\expandafter\ifx\csname bibfnamefont\endcsname\relax
  \def\bibfnamefont#1{#1}\fi
\expandafter\ifx\csname citenamefont\endcsname\relax
  \def\citenamefont#1{#1}\fi
\expandafter\ifx\csname url\endcsname\relax
  \def\url#1{\texttt{#1}}\fi
\expandafter\ifx\csname urlprefix\endcsname\relax\def\urlprefix{URL }\fi
\providecommand{\bibinfo}[2]{#2}
\providecommand{\eprint}[2][]{\url{#2}}

\bibitem[{\citenamefont{Bekenstein}(1972)}]{Bekenstein:1972ny}
\bibinfo{author}{\bibfnamefont{J.~D.} \bibnamefont{Bekenstein}},
  \bibinfo{journal}{Phys. Rev. Lett.} \textbf{\bibinfo{volume}{28}},
  \bibinfo{pages}{452} (\bibinfo{year}{1972}),
  \urlprefix\url{https://link.aps.org/doi/10.1103/PhysRevLett.28.452}.

\bibitem[{\citenamefont{Herdeiro and Radu}(2014)}]{Herdeiro:2014goa}
\bibinfo{author}{\bibfnamefont{C.~A.~R.} \bibnamefont{Herdeiro}}
  \bibnamefont{and} \bibinfo{author}{\bibfnamefont{E.}~\bibnamefont{Radu}},
  \bibinfo{journal}{Phys. Rev. Lett.} \textbf{\bibinfo{volume}{112}},
  \bibinfo{pages}{221101} (\bibinfo{year}{2014}),
  \urlprefix\url{https://link.aps.org/doi/10.1103/PhysRevLett.112.221101}.

\bibitem[{\citenamefont{Hong et~al.}(2020)\citenamefont{Hong, Suzuki, and
  Yamada}}]{Hong:2020miv}
\bibinfo{author}{\bibfnamefont{J.-P.} \bibnamefont{Hong}},
  \bibinfo{author}{\bibfnamefont{M.}~\bibnamefont{Suzuki}}, \bibnamefont{and}
  \bibinfo{author}{\bibfnamefont{M.}~\bibnamefont{Yamada}},
  \bibinfo{journal}{Phys. Rev. Lett.} \textbf{\bibinfo{volume}{125}},
  \bibinfo{pages}{111104} (\bibinfo{year}{2020}),
  \urlprefix\url{https://link.aps.org/doi/10.1103/PhysRevLett.125.111104}.

\bibitem[{\citenamefont{Huang and Liu}(2016)}]{Huang:2016qnk}
\bibinfo{author}{\bibfnamefont{Y.}~\bibnamefont{Huang}} \bibnamefont{and}
  \bibinfo{author}{\bibfnamefont{D.-J.} \bibnamefont{Liu}},
  \bibinfo{journal}{Phys. Rev. D} \textbf{\bibinfo{volume}{94}},
  \bibinfo{pages}{064030} (\bibinfo{year}{2016}),
  \urlprefix\url{https://link.aps.org/doi/10.1103/PhysRevD.94.064030}.

\bibitem[{\citenamefont{Huang et~al.}(2017{\natexlab{a}})\citenamefont{Huang,
  Liu, Zhai, and Li}}]{Huang:2017whw}
\bibinfo{author}{\bibfnamefont{Y.}~\bibnamefont{Huang}},
  \bibinfo{author}{\bibfnamefont{D.-J.} \bibnamefont{Liu}},
  \bibinfo{author}{\bibfnamefont{X.-H.} \bibnamefont{Zhai}}, \bibnamefont{and}
  \bibinfo{author}{\bibfnamefont{X.-Z.} \bibnamefont{Li}},
  \bibinfo{journal}{Classical and Quantum Gravity}
  \textbf{\bibinfo{volume}{34}}, \bibinfo{pages}{155002}
  (\bibinfo{year}{2017}{\natexlab{a}}),
  \urlprefix\url{https://doi.org/10.1088/1361-6382/aa7964}.

\bibitem[{\citenamefont{Hod}(2012)}]{Hod:2012px}
\bibinfo{author}{\bibfnamefont{S.}~\bibnamefont{Hod}}, \bibinfo{journal}{Phys.
  Rev. D} \textbf{\bibinfo{volume}{86}}, \bibinfo{pages}{104026}
  (\bibinfo{year}{2012}),
  \urlprefix\url{https://link.aps.org/doi/10.1103/PhysRevD.86.104026}.

\bibitem[{\citenamefont{Bernard}(2016)}]{Bernard:2016wqo}
\bibinfo{author}{\bibfnamefont{C.}~\bibnamefont{Bernard}},
  \bibinfo{journal}{Phys. Rev. D} \textbf{\bibinfo{volume}{94}},
  \bibinfo{pages}{085007} (\bibinfo{year}{2016}),
  \urlprefix\url{https://link.aps.org/doi/10.1103/PhysRevD.94.085007}.

\bibitem[{\citenamefont{Gibbons and ichi Maeda}(1988)}]{Gibbons:1987ps}
\bibinfo{author}{\bibfnamefont{G.}~\bibnamefont{Gibbons}} \bibnamefont{and}
  \bibinfo{author}{\bibfnamefont{K.}~\bibnamefont{ichi Maeda}},
  \bibinfo{journal}{Nuclear Physics B} \textbf{\bibinfo{volume}{298}},
  \bibinfo{pages}{741} (\bibinfo{year}{1988}), ISSN \bibinfo{issn}{0550-3213},
  \urlprefix\url{https://www.sciencedirect.com/science/article/pii/0550321388900065}.

\bibitem[{\citenamefont{Garfinkle et~al.}(1991)\citenamefont{Garfinkle,
  Horowitz, and Strominger}}]{Garfinkle:1990qj}
\bibinfo{author}{\bibfnamefont{D.}~\bibnamefont{Garfinkle}},
  \bibinfo{author}{\bibfnamefont{G.~T.} \bibnamefont{Horowitz}},
  \bibnamefont{and}
  \bibinfo{author}{\bibfnamefont{A.}~\bibnamefont{Strominger}},
  \bibinfo{journal}{Phys. Rev. D} \textbf{\bibinfo{volume}{43}},
  \bibinfo{pages}{3140} (\bibinfo{year}{1991}),
  \urlprefix\url{https://link.aps.org/doi/10.1103/PhysRevD.43.3140}.

\bibitem[{\citenamefont{Brill and Wheeler}(1957)}]{RevModPhys.29.465}
\bibinfo{author}{\bibfnamefont{D.~R.} \bibnamefont{Brill}} \bibnamefont{and}
  \bibinfo{author}{\bibfnamefont{J.~A.} \bibnamefont{Wheeler}},
  \bibinfo{journal}{Rev. Mod. Phys.} \textbf{\bibinfo{volume}{29}},
  \bibinfo{pages}{465} (\bibinfo{year}{1957}),
  \urlprefix\url{https://link.aps.org/doi/10.1103/RevModPhys.29.465}.

\bibitem[{\citenamefont{Lasenby et~al.}(2005)\citenamefont{Lasenby, Doran,
  Pritchard, Caceres, and Dolan}}]{Lasenby:2002mc}
\bibinfo{author}{\bibfnamefont{A.}~\bibnamefont{Lasenby}},
  \bibinfo{author}{\bibfnamefont{C.}~\bibnamefont{Doran}},
  \bibinfo{author}{\bibfnamefont{J.}~\bibnamefont{Pritchard}},
  \bibinfo{author}{\bibfnamefont{A.}~\bibnamefont{Caceres}}, \bibnamefont{and}
  \bibinfo{author}{\bibfnamefont{S.}~\bibnamefont{Dolan}},
  \bibinfo{journal}{Phys. Rev. D} \textbf{\bibinfo{volume}{72}},
  \bibinfo{pages}{105014} (\bibinfo{year}{2005}),
  \urlprefix\url{https://link.aps.org/doi/10.1103/PhysRevD.72.105014}.

\bibitem[{\citenamefont{Huang and Zhang}(2021)}]{Huang:2020pga}
\bibinfo{author}{\bibfnamefont{Y.}~\bibnamefont{Huang}} \bibnamefont{and}
  \bibinfo{author}{\bibfnamefont{H.}~\bibnamefont{Zhang}},
  \bibinfo{journal}{Phys. Rev. D} \textbf{\bibinfo{volume}{103}},
  \bibinfo{pages}{044062} (\bibinfo{year}{2021}),
  \urlprefix\url{https://link.aps.org/doi/10.1103/PhysRevD.103.044062}.

\bibitem[{\citenamefont{Dias et~al.}(2016)\citenamefont{Dias, Santos, and
  Way}}]{Dias:2015nua}
\bibinfo{author}{\bibfnamefont{O.~J.~C.} \bibnamefont{Dias}},
  \bibinfo{author}{\bibfnamefont{J.~E.} \bibnamefont{Santos}},
  \bibnamefont{and} \bibinfo{author}{\bibfnamefont{B.}~\bibnamefont{Way}},
  \bibinfo{journal}{Class. Quant. Grav.} \textbf{\bibinfo{volume}{33}},
  \bibinfo{pages}{133001} (\bibinfo{year}{2016}), \eprint{1510.02804}.

\bibitem[{\citenamefont{Ternov et~al.}(1980)\citenamefont{Ternov, Gaina, and
  Chizhov}}]{Ternov:1980st}
\bibinfo{author}{\bibfnamefont{I.~M.} \bibnamefont{Ternov}},
  \bibinfo{author}{\bibfnamefont{A.~B.} \bibnamefont{Gaina}}, \bibnamefont{and}
  \bibinfo{author}{\bibfnamefont{G.~A.} \bibnamefont{Chizhov}},
  \bibinfo{journal}{Sov. Phys. J.} \textbf{\bibinfo{volume}{23}},
  \bibinfo{pages}{695} (\bibinfo{year}{1980}).

\bibitem[{\citenamefont{Huang et~al.}(2017{\natexlab{b}})\citenamefont{Huang,
  Liu, Zhai, and Li}}]{Huang:2017nho}
\bibinfo{author}{\bibfnamefont{Y.}~\bibnamefont{Huang}},
  \bibinfo{author}{\bibfnamefont{D.-J.} \bibnamefont{Liu}},
  \bibinfo{author}{\bibfnamefont{X.-h.} \bibnamefont{Zhai}}, \bibnamefont{and}
  \bibinfo{author}{\bibfnamefont{X.-z.} \bibnamefont{Li}},
  \bibinfo{journal}{Phys. Rev. D} \textbf{\bibinfo{volume}{96}},
  \bibinfo{pages}{065002} (\bibinfo{year}{2017}{\natexlab{b}}),
  \urlprefix\url{https://link.aps.org/doi/10.1103/PhysRevD.96.065002}.

\bibitem[{\citenamefont{Holzhey and Wilczek}(1992)}]{Holzhey:1991bx}
\bibinfo{author}{\bibfnamefont{C.~F.} \bibnamefont{Holzhey}} \bibnamefont{and}
  \bibinfo{author}{\bibfnamefont{F.}~\bibnamefont{Wilczek}},
  \bibinfo{journal}{Nuclear Physics B} \textbf{\bibinfo{volume}{380}},
  \bibinfo{pages}{447} (\bibinfo{year}{1992}), ISSN \bibinfo{issn}{0550-3213},
  \urlprefix\url{https://www.sciencedirect.com/science/article/pii/0550321392902549}.

\end{thebibliography}

\end{document}